\documentclass{article}
\usepackage[T1]{fontenc}
\usepackage[english]{babel}
\usepackage{amsfonts,amssymb, amsmath}

\begin{document}

\title{On the Lie noncommutative integrability}
\author{A.\,V.~Tsiganov\\
\it\small St. Petersburg State University, St. Petersburg, Russia;\\
\it\small Beijing Institute of Mathematical Sciences and Applications, Beijing, China;\\
\it\small e-mail: andrey.tsiganov@gmail.com
}
\date{}
\maketitle

\begin{abstract}
The Lie theory of non-commutative integrability is used to reconstruct some integrable systems of ordinary differential equations in $\mathbb R^3$.  The  Darboux–Brioschi–Halphen system is an example of a Lie-integrable system associated with the simple Lie algebra $sl(2,\mathbb{R})$. Other examples are related to  solvable three-dimensional real Lie algebras.
 \end{abstract}

\newtheorem{prp}{Proposition}

\section{Introduction}
\setcounter{equation}{0}

A system of ordinary differential equations
\begin{equation}\label{ode-1}
\frac{dx_1}{\xi_1}=\frac{dx_2}{\xi_2}=\cdots=\frac{dx_n}{\xi_n}
\end{equation}
can be completely integrable, incompletely integrable or non-integrable; see the classical textbooks  \cite{im69,lie-book,lie-book2, for90,gour91,car}. For a completely integrable system (\ref{ode-1}) there exist single-valued solutions in the neighborhood of any given point that satisfies the specified conditions. These local solutions $x_k$ depend on certain prescribed initial values of $x_k=x'_k$ at the given point, which have to be interpreted as the integration constants.

By inverting the relations between $x_k$ and $x'_k$ we obtain integral functions $f_k(x_1,\ldots,x_n)$ which are solutions to a first-order linear partial differential equation
\[
X(f) =\sum \xi_{i}(x_1,\ldots,x_n)\frac{\partial f}{\partial x_i}\,,
\]
see Jacobi's original paper \cite{jac-27} and Lie's textbook \cite{lie-book}. The integral functions can be local, multi-valued and transcendental functions. There are a few criteria for complete integrability and the corresponding constructions of completely integrable systems that do not require the exact calculation of integral functions. Various such constructions were proposed by Grassmann, Hamburger, Deahna, Clebsch, Natani, Mayer, Lie, Frobenius, Darboux and Cartan, see \cite{for90,car}.

Our aim is to discuss Lie's classical constructions of completely integrable systems based on the groups of infinitesimal transformations and homogeneous function groups. In modern terms 
Lie’s theory of symmetry groups of differential equations involves realisations of Lie algebras in the space of vector fields, in the space of functions, in jet spaces, etc. \cite{olver}.  

\subsection{Infinitesimal transformations preserving Pfaff's equations}
The problem, which is known as Pfaff's problem, originated in the theory of first-order partial differential equations, which was developed by Lagrange, see, e.g., \cite{lag81,lag85}. For a partial differential equation 
\[
F(x_1,\ldots,x_n,z, p_1,\ldots, p_n)=0\,,\qquad p_i=\frac{\partial z}{\partial x_i}
\] 
where $z$ is singled out as the dependent variable, Lagrange's aim was to obtain a 
"complete solution" depending on $n$ arbitrary constants 
\[
z = \varphi(x_1,\ldots,x_n;C_1,\ldots,C_n)
\]
Here by “obtain a solution” Lagrange means to show how such a solution can be obtained
by means of solutions to one or more systems of ordinary differential equations:
\par\noindent
“{\em the art of the calculus of
partial derivatives is known to consist in nothing more than reducing this calculus to
that of ordinary derivatives, and a partial differential equation is regarded as integrated
when its integral depends on nothing more than that of one or more ordinary differential
equations}”, see page 625 in \cite{lag81}.

For linear first order equations, Lagrange showed how to do this for any number of
variables \cite{lag85}. The integration of nonlinear equations with
$n > 2$ was first achieved  by Pfaff in 1815 \cite{pfaff}. Pfaff studied
the more general problem of “integrating,” in a sense to be discussed, a total differential equation
\[
\omega= a_1(x)dx_1 +\cdots+a_n(x)dx_n = 0 
\]
in variables $x = (x_1,\ldots , x_n)$.

According to Pfaff “in general” a change of variables 
$x_i =\alpha_i(y_1,\ldots , y_n)$
exists such that 
\[
\omega= a_1(x)dx_1 +\cdots+a_n(x)dx_n = 0 
\]
transforms into an expression involving $m$ differentials
\[\omega = b_1(y)dy_1 +\cdots+b_m(y)dy_m,\]
where $m = n/2$ if $n$ is even and $m = (n + 1)/2$ if $n$ is odd \cite{pfaff}. In 1827 Jacobi proved Pfaff’s theorem in the case of $n = 2m$ variables, the case that was relevant to its application to partial differential equations \cite{jac-27}. 

In 1872 Lie came away from the geometrical period of research with more than the
notion of a continuous group of transformations and an appreciation of its relevance
to geometry.  He had come to believe that it was possible to develop something analogous to Galois' theory of algebraic equations for the study of differential equations. 

The original Lie works \cite{lie-00, lie-1873a,lie-1873c,lie-0,lie-1,lie-1874b} examine groups of infinitesimal transformations that preserve Pfaff’s equations. Then he proposed a new interpretation of solutions to complete systems and attempted to determine all systems of differential equations that admit a given group of infinitesimal transformations, see the corresponding chapters in his textbook \cite{lie-book}. Therefore, it is important to describe the notion of infinitesimal transformations as used by Lie in 1872–73.

Following \cite{lie-00} consider a first order partial differential equation
\begin{equation}\label{lag-pde} F(x_1, x_2, z, p_1, p_2) = 0,\qquad p_k=\frac{\partial z}{\partial x_k}\,.
\end{equation} 
This equation admits a transformation
\[(x_1, x_2, z)\to (y_1, y_2, \zeta)\]
if every solution surface $\varphi(x_1, x_2, z) = 0$ of the equation
is transformed into a solution surface \cite{lie-00}. This is what Lie usually meant when he said that a differential equation or a system of differential equations admits a transformation: the transformation turns solutions into solutions. An equation admits a group of transformations if it admits all the transformations that comprise it. 

In 1872 Lie wrote \textit{ "I have succeeded in broadening my work on partial [differential] equations with
infinitesimal transformations in various directions and in particular have also
extended it to the Pfaff Problem and to systems of ordinary differential equations.
I consider both commutative transformations as well as those which form a group"}, see memoir \cite{lie-00}.

Later Lie formulated the following noncommutative integrability problem:\\
\textit{"Suppose a system of ordinary differential equations or, equivalently,
a system of independent linear, homogeneous partial differential equations
\[
X_i(f)=\sum_{j=1}^n \xi_{ij}(x)\frac{\partial f}{\partial x_j}=0\,,\qquad i=1,\cdots,r\leq n\,,
\]
is known to admit infinitesimal transformations (which perhaps commute or form
a group). What does this information imply about the integration of the system?"}



The title of Lie’s paper \cite{lie77}, "Theory of the Pfaffian Problem I", suggested a sequel, and Lie had indeed entertained the idea of a substantial second part containing an extension to Pfaff's general problem of his theory of contact transformations, first-order partial differential equations, and transformation groups. This part never materialised. In 1883 Lie explained to Mayer that
  \textit{“Because of Frobenius I have lost interest in the problem of Pfaff . . . . I have already written too much that goes unread”} see page 714 in \cite{engel}. 
  
  Since the Berlin school was one of the most prestigious and influential centres for mathematics in the 1870s, Lie's remarks probably reflect the seeming hopelessness of competing with the way the Berlin school approached the Pfaff problem, as presented by Frobenius \cite{frob}, see comments in \cite{for90}.


The significance of Lie's "new interpretation of complete solutions of differential systems" was that, from the knowledge that the partial differential equation admits a group of known transformations, one is able to obtain information about its integration, about what is involved to effect its integration. 

\subsection{Function groups} 
The Lie theory of function groups represents a precursor of the theory of Poisson structures associated with partial differential equations and the corresponding ordinary differential equations \cite{lie-1873a,lie-1873c}. In presenting his theory of homogeneous function groups, Lie expressly referred to \S14 of \cite{im69}.

According to Lie, infinitesimal transformations $X_i(f)$, $i = 1,\cdots, r$ determine an $r$-dimensional group of transformations of the Pfaffian equation $\omega=0$ if they satisfy 
\begin{equation}\label{lie-group}
X_i(X_j(f))-X_j(X_i(f))= \sum_{k=1}^r c_{ij}^kX_k(f).
\end{equation}
Here, the $c_{ij}^k$ satisfy the Jacobi identities and may be either constants or functions of the first integrals, see \cite{lie-book,lie-book2} and Cartan's textbook  \cite{car}. If the $c_{ij}^k$ are functions of the first integrals, we have Lie's theory of characteristic strips, which is beyond the scope of this note.

In the special case of the Pfaffian equation
\begin{equation}\label{spec-pfaff}
\omega= dz-p_1dx_1-\cdots-p_ndx_n= 0
\end{equation}
each infinitesimal transformation $X_i(f)$ has its characteristic function
$H_i(x,p)$ so that \[X_i(f) = (H_i,f)\qquad\mbox{for all}\qquad f = f ( x , p )\,.\] 
In the Lagrange-Pfaff theory the equation $\omega=0$ (\ref{spec-pfaff}) is  another form of the partial differential equation (\ref{lag-pde}), and the infinitesimal transformation $X_i$ is a symmetry of the equations (\ref{lag-pde}, \ref{spec-pfaff}). This theorem was first published in \cite{lie-1874b}. An analogous theorem for generic Pfaffian equations and generic infinitesimal transformations preserving solutions may be found in the book \cite{lie-book}.

Here $(H_i,f)$ is a standard Poisson bracket on the symplectic manifold $\mathbb R^{2n}$ but, of course, Lie's papers do not contain the modern terms "symplectic manifold", "Hamiltonian vector field" for (\ref{spec-pfaff}) or "differential 1-form on $\mathbb{R}^{n+1}$" for $\omega$. In fact, Lie used the term "manifold" to denote a totality of "elements" of the form $x =(x_1,\ldots,x_n)$. The modern conception of a manifold was introduced by  Weyl in the 20th century, see \cite{sch80}. 

Then Lie combines this theorem with (\ref{lie-group}) and proves that for all $f$
\[((H_i,H_j),f) = \sum_{k=1}^r c_{ij}^kX_k(f) = \sum_{k=1}^r c_{ij}^k(H_k,f)\,,\]
so that the characteristic functions $H_i(x, p)$ satisfy
\begin{equation}\label{h-group}
 (H_i,H_j)= \sum_{k=1}^r c_{ij}^kH_k\,,
 \end{equation}
which is an exact analog of (\ref{lie-group}), see memoir \cite{lie-1874b}. 

As Lie  himself explained a decade
later on pages 140--141 in \cite{lie-1884a}: \textit{"In my investigations of line and sphere geometry in the years 1870--71
many similar integration methods were applied with some success
to special geometrical problems. Soon however these investigations took on an
entirely different significance. I discovered that the integration theory of first order
partial differential equations founded by Lagrange, Pfaff, Cauchy, Jacobi and
their successors could be conceived in a natural way as a transformation theory
of these equations ... . Every known group of  transformations which
took a first order partial differential equation  into itself supplied, corresponding
to its infinitesimal transformations, a certain number of integrals $u_1,\ldots, u_n$
of the associated simultaneous system  which satisfied pairwise
relations of the form $(u_i, u_k) =f_k(u_1,\ldots,u_n)$."}.

The results obtained by Lie and Mayer on the noncommutative integration using function groups (\ref{h-group}) are referred to in Steklov's works \cite{st1,st2} as a special case of the generalized Jacobi theorem. In \cite{fom}, Mishchenko and Fomenko partially reinstated this generalised Jacobi theorem.

\subsection{On the Jacobi theory of multipliers}
In \cite{lie-0,lie-1}, see also  \S 67 of his book \cite{lie-book},  Lie considers simply transitive groups  which contain $r=n$ independent infinitesimal transformations 
\[
X_i(f) =\sum \xi_{i,j}(x_1,\ldots,x_n)\partial_j (f)
\]
for which the determinant
\[
\Delta=\left\| 
\begin{array}{ccc}
\xi_{1,1}(x)& &\xi_{1,n}(x)\\
\vdots& &\vdots\\
\xi_{n,1}(x)&\quad\cdots \quad &\xi_{n,n}(x)\\
\end{array}
\right\|\,,
\]
does not vanish identically. In this case, the function
\begin{equation*}
M=\Delta^{-1}
\end{equation*}
satisfies the Jacobi equations \cite{jac}
\[
\mbox{div} (MX_i)=0\,,\qquad i=1,\ldots,n,
\]
for all the infinitesimal transformations $X_1,\ldots,X_n$.
So, for Lie's simple transitive groups  we have $n$ independent complete partial differential equations
\[
X_i(f)=0
\]
with the common multiplier $M=\Delta^{-1}$ \cite{lie-0,lie-1}. For other groups of infinitesimal transformations complete partial differential equations $X_i(f)=0$ have different multipliers, see details in textbook \cite{lie-book}. 

Unfortunately, the term "Jacobi multiplier" does not appear in Lie's textbook \cite{lie-book}. Modern theory of the Jacobi multiplier may be found in Cartan's textbook \cite{car}, see also \cite{koz}.

If an infinitesimal transformation group is integrable we can obtain the corresponding complete solution by using rectifying coordinates \cite{lie-book,lie-book2}. In modern literature authors use the term "solvable Lie algebras" instead of Lie's term "integrable infinitesimal transformation group".

In the next section we consider representations of the low-dimensional groups of infinitesimal transformations and the corresponding invariant differential equations. Following Lie, we will primarily consider homogeneous vector fields and homogeneous differential systems. It means that the group of infinitesimal transformations contains the Euler vector field
 that generates scalar multiplication. On a Euclidean space it points radially outward from the origin, stretching or scaling space.
 
\section{Solvable real Lie algebras}
In \cite{lie-book,lie-book2}, Lie studied so-called integrable groups of infinitesimal transformations. In modern terms, this is the study of solvable Lie algebras.

If $\mathfrak g$ is a real solvable Lie algebra, $\dim\mathfrak g\leq 3$, then $\mathfrak g$ is isomorphic
to one and only one of the following Lie algebras: $\mathbb R$, $\mathbb R^2$, $\mathfrak{aff}(\mathbb R)$, $\mathbb R^3$, $\mathfrak h_3$, $\mathfrak r_3$, $\mathfrak r_{3,\lambda},\, |\lambda|\leq 1$ and $\mathfrak r'_{3,\lambda},\,\lambda\geq 0$  with nontrivial brackets between generators:
\begin{align*} 
\mathfrak{aff}(\mathbb R):\qquad &[e_1,e_2]=e_2;\\ 
\mathfrak{h}_3:\qquad &[e_1,e_2] = e_3; \\ 
\mathfrak{r}_3: \qquad &[e_1,e_2]=e_2,\quad [e_1,e_3]=e_2+e_3;\\ 
\mathfrak{r}_{3,\lambda}:\qquad &[e_1,e_2]=e_2,\quad   [e_1,e_3]= \lambda e_3;\\ 
\mathfrak{r}'_{3,\lambda}:\qquad &[e_1,e_2]=\lambda e_2 - e_3,  \quad [e_1,e_3]=e_2 +\lambda e_3;
\end{align*}
see details in \cite{gov}. Here $\mathfrak{aff}(\mathbb R)$ is the two-dimensional non-abelian Lie algebra of the group of affine motions of the real line, $\mathfrak h_3$ is the three-dimensional Heisenberg algebra, $\mathfrak r_{3,-1}$ is the Lie algebra $\mathfrak{e(1, 1)}$ of the group of rigid motions
of Minkowski 2-space, $\mathfrak r_{3,0} = \mathbb R \times \mathfrak{aff}(\mathbb R)$ and $\mathfrak r_{3,1}$ is the Lie algebra of the solvable group which acts simply and transitively on the real hyperbolic space $\mathbb RH^3$.
Also $\mathfrak r'_{3,0}$ is the Lie algebra $\mathfrak{e(2)}$ of the group of rigid motions of Euclidean 2-space. 

Below we consider some  realizations of these real algebras and the corresponding completely integrable differential equations. All the parameters are assumed to be generic real numbers, so that none of the denominators appearing in the displayed formulae vanishes; degenerate cases require separate limiting
formulae.

\subsection{Solvable algebra $\mathfrak{aff}(\mathbb R)$}
In fact, all the $\mathfrak{aff}(\mathbb R)$-invariant vector fields were studied by  Euler in \S477 of his textbook  \cite{eul} where he considered a completely integrable differential system in $\mathbb R^2$
\begin{equation}\label{e-pfaff}
\omega=P(x, y)dx+Q(x, y)dy=0\,,
\end{equation}
where $P(x, y)$ and $Q(x, y)$ are homogeneous functions  of degree $\kappa$, which satisfy Euler's equations 
\[
x\partial_x P+y\partial_y P=\kappa P\qquad\mbox{and}\qquad
x\partial_x Q+y\partial_y Q=\kappa Q\,,\qquad \kappa\in\mathbb R\,.
\]
The corresponding  vector field
\[
X=Q(x,y)\partial_x -P(x,y)\partial_y\,,
\]
has the following Lie bracket
\[[X,Y]=(\kappa-1)X\]
with the Euler vector field $Y=x\partial_x+y\partial_y$, which is a symmetry of the one-dimensional distribution generated by $X$. So, Lie's group of infinitesimal transformations that preserve Pfaff's equation (\ref{e-pfaff})  is isomorphic to $\mathfrak{aff}(\mathbb R)$. 

Lie's determinant 
\[\Delta=\left\| 
\begin{array}{cc}
Q&-P\\
x&y\\
\end{array}
\right\|\neq 0\,,\]
defines Jacobi's multiplier 
\[
M(x,y)=\Delta^{-1}=\frac{1}{x P+yQ}\neq 0\,, 
\]
for the vector field $X$ but not for $Y$.

This multiplier $M$ together with the first integral $\varphi(f)$
\[
f(x,y)=
\int  \frac{P}{xP  +  yQ}dx+\int \frac{Q}{x P+yQ} dy+
\iint
\frac{(1-\kappa)QP - xP\partial_xQ-yQ\partial_y P }{( xP  +  yQ)^2}dxdy\,.
\]
were found by Euler from the equations
\[
MP=\frac{\partial \varphi}{\partial x} \qquad\mbox{and}\qquad MQ=\frac{\partial \varphi}{\partial y}\,.
\]
According to Euler, $\varphi(f)$ can be a polynomial, rational or transcendental function  \cite{eul}. 

\subsection{Solvable algebra $\mathfrak r_3$}
Let us consider a solvable algebra with the following Lie brackets 
\begin{equation}\label{t3}
[e_1,e_2]=e_2\,,\qquad [e_1,e_3]=e_2+e_3\,,\qquad [e_2,e_3]=0\,.
\end{equation}
Let us consider a realisation of the three-dimensional real solvable algebra in the space of vector fields on $\mathbb R^3$ so that 
\begin{equation}\label{e12-sol}
e_1=\frac{1}{\kappa-1}(x_1\partial_1+x_2\partial_2+x_3\partial_3)\,\qquad
e_2=a_1x_1^\kappa\partial_1+a_2x_2^\kappa\partial_2+a_3x_3^\kappa\partial_3\,,
\end{equation}
and 
\begin{equation}\label{e3-sol}
e_3=\phi_1(x_1,x_2,x_3)\partial_1+\phi_2(x_1,x_2,x_3)\partial_2+\phi_3(x_1,x_2,x_3)\partial_3
\end{equation}
where $\kappa\neq 1$ and functions $\phi_k(x)$ are solutions of the PDEs which follow from the Lie brackets (\ref{t3}). 

Invariants of Euler's vector field $e_1$ are homogeneous functions of degree zero, whereas invariants of $e_2$ are functions on
\[
d_{ij}=a_ix_j^{1-\kappa}-a_jx_i^{1-\kappa}\,.
\]
We can use these invariants to construct functions $f_k$ and the corresponding ODEs following Lie's textbooks \cite{lie-book, lie-book2}.
\begin{prp}
Solving the corresponding partial differential equations  we obtain 
\[ 
e_3=x_1^\kappa\,\Bigl(-a_1\ln D +g_1(d_{ij})\Bigr)\partial_1
   +x_2^\kappa\,\Bigl(-a_2\ln D +g_2(d_{ij})\Bigr)\partial_2
   +x_3^\kappa\,\Bigl(-a_3\ln D +g_3(d_{ij})\Bigr)\partial_3\,,
\]
where \[D=\sum c^{ij}d_{ij}\,,\qquad c^{ij}\in\mathbb R\,,
\]
and $g_k(d_{ij})\equiv g_k(d_{12},d_{23},d_{3,1})$ are homogeneous functions of zero degree on $x$ which depend only on the invariants $d_{ij}$ of $e_2$.
As an example we can take
\[
g_k(d_{ij})=G_k\left(\frac{d_{13}}{d_{12}},\frac{d_{23}}{d_{12}}\right)\,.
\]
\end{prp}
The proof consists of directly solving the corresponding partial differential equations.

According to the Jacobi-Clebsch-Deahna-Frobenius-Lie theorem we have a system of completely integrable differential equations of the form
\[
\dot{x}_k=\phi_k\equiv x_k^\kappa\,\Bigl(-a_k\ln D + G_k\left(\frac{d_{13}}{d_{12}},\frac{d_{23}}{d_{12}}\right))\Bigr)\,,\qquad k=1,2,3.
\]
The corresponding  Lie's determinant
\[
\Delta=\left\| 
\begin{array}{ccc}
\beta x_1&\beta x_2 &\beta x_3\\
a_1x_1^\kappa&a_2x_2^\kappa&a_3x_3^\kappa\\
\phi_1&\phi_2&\phi_3\\
\end{array}
\right\|\,,\qquad \beta=\frac{1}{\kappa-1}\,,
\]
defines Jacobi's multiplier $M=\Delta^{-1}$ for the flows associated with vector fields $e_2$ and $e_3$ at $\Delta\neq 0$.

This example demonstrates the construction of a locally complete non-algebraic differential system in the framework of Lie's noncommutative integrability theory.  

\subsection{Solvable algebra $\mathfrak r_{3,\lambda}$}
Let us consider a solvable algebra with the following Lie brackets 
\begin{equation}\label{t3l}
[e_1,e_2]=e_2\,,\qquad [e_1,e_3]=\lambda e_3\,,\qquad [e_2,e_3]=0\,.
\end{equation}
\begin{prp}
Substituting $e_{1,2}$ (\ref{e12-sol}) and $e_3$ (\ref{e3-sol}) into  (\ref{t3l}) and solving the corresponding partial differential equations we obtain 
\[ 
e_3=x_1^\kappa\,g_1(d_{ij})\partial_1+x_2^\kappa\,g_2(d_{ij})\partial_2+x_3^\kappa\,g_3(d_{ij})\partial_3\,,
\]
where $g_k(d)$ are homogeneous functions of degree $(1-\lambda)(1-\kappa)$ on $x$ which depend only on the invariants $d_{ij}$ of $e_2$. As an example we can take
\[
g_k(d)=d_{1,2}^{1-\lambda}\,G_k\left(\frac{d_{13}}{d_{12}},\frac{d_{23}}{d_{12}}\right)\,.
\]
\end{prp}
The proof consists of directly solving the corresponding partial differential equations.

According to the Jacobi-Clebsch-Deahna-Frobenius-Lie theorem we have a system of completely integrable differential equations of the form
\[
\dot{x}_k=\phi_k\equiv x_k^\kappa\,d_{1,2}^{1-\lambda}\,G_k\left(\frac{d_{13}}{d_{12}}\right)\,,\qquad k=1,2,3,
\]
for $|\lambda|<1$. 

The Lie determinant is equal to
\[
\Delta=\left\| 
\begin{array}{ccc}
\beta x_1&\beta x_2 &\beta x_3\\
a_1x_1^\kappa&a_2x_2^\kappa&a_3x_3^\kappa\\
\phi_1&\phi_2&\phi_3\\
\end{array}
\right\|\,,\qquad \qquad \beta=\frac{1}{\kappa-1}\,,
\]
and Jacobi's multiplier $M=\Delta^{-1}$ at $\Delta\neq 0$
for the flows associated with $e_2$ and $e_3$.

Since $|\lambda|<1$ this example also demonstrates the construction of a locally complete differential system in the framework of Lie's noncommutative integrability theory. 

\subsection{Solvable algebra $\mathfrak{r}_{3,0}$}
Some examples of the completely integrable May-Leonard, Lotka-Volterra and Lorenz equations  were obtained using various  realisations of $\mathfrak{r}_{3,0}$ in \cite{sw1,sw2,os90}. The corresponding invariant differential systems have one, two or no single-valued first integrals. 

Let us present one example of the Lotka-Volterra system associated with the vector field 
\[e_2=x_1(-x_2 + x_3)\partial_1+x_2(x_1 + x_3)\partial_2+x_3(x_1+x_2)\partial_3\,,\]
Euler vector field
\[e_1=x_1\partial_1+x_2\partial_2+x_3\partial_3\,.\]
It is Example 4 from Table 1 in \cite{sw1}, where the authors study two vector fields in the framework of the Frobenius theorem. 
 
We want to construct a realisation of the solvable Lie algebra $\mathfrak{r}_{3,0}$ which was omitted in \cite{sw1}, i.e. to construct a third vector field
\[
e_3=\phi_1(x)\partial_1+\phi_2(x)\partial_2+\phi_3(x)\partial_3
\]
that allows us to apply Lie's theorem if
\begin{equation}\label{e2}
[e_1,e_2]= e_2\,,\qquad [e_1,e_3]= 0\,,\qquad [e_2,e_3]=0\,.
\end{equation}
One of the particular solutions of the corresponding PDEs has the following form
\[
\phi_1=\frac{x_1x_2}{x_1 + x_2}\,,\qquad 
\phi_2=x_2^2\left(\frac{1}{x_1}+\frac{1}{x_1+x_2}\right)\,,\qquad
\phi_3=\frac{x_2(x_1 + x_2)}{x_1}\,.
\] 
In this case we can directly prove complete integrability without calculation of the first integral as in \cite{sw1}.

In \cite{sw1,sw2,os90} the authors construct realisations of $\mathfrak r_{3,0}$ starting with known completely integrable systems. We will demonstrate that we can use the inverse approach to construct new completely integrable systems starting with simple realisations of the same algebra.

As a first example we substitute Euler's vector field
\[e_1=x_1\partial_1+x_2\partial_2+x_3\partial_3\,,\]
vector fields
\[
e_2=\phi_1(x)\partial_1+\phi_2(x)\partial_2+\phi_3(x)\partial_3
\]
and
\[
e_3=x_1 (a\partial_1+b\partial_2+c\partial_3)\,,\qquad abc\neq 0\,,
\]
into (\ref{e2}). In this case, the functions $\phi_k$ are obtained by solving the PDEs associated with the brackets (\ref{e2}). One of the solutions is
\[
e_2= (x,{\mathbf A}x)\partial_1+(x,{\mathbf B}x)\partial_2+(x,{\mathbf C}x)\partial_3\,,\qquad x=(x_1,x_2,x_3)\,,
\]
whose entries are second-order polynomials in $x_k$. Here
\begin{align*}
\mathbf A=&\left(
            \begin{array}{ccc}
              -2 (bA_1b + cA_2c) & aA_1 & aA_2 \\
              aA_1 & 0 & 0 \\
              aA_2 & 0 & 0 \\
            \end{array}
          \right)
          \\
          \\
\mathbf B=&\left(
            \begin{array}{ccc}
              \dfrac{b^2(2A_1 - B_1) + 2bc(A_2 - B_2) - c^2B_3}{a} & b(A_1 - B_1) - cB_2 & b(A_2 - B_2) - cB_3\\
              b(A_1 - B_1) - cB_2 & aB_1 & aB_2 \\
              b(A_2 - B_2) - cB_3 &a B_2 & aB_3 \\
            \end{array}
          \right)
\end{align*}
and
\[
\mathbf C=\left(
             \begin{array}{ccc}
             \dfrac{c^2(2A_2 - C_3) + 2bc(A_1 - C_2) - b^2C_1}{a}   & c(A_1- C_2) - bC_1 & c(A_2 - C_3) - bC_2 \\
               c(A_1- C_2) - bC_1 & aC_1 & aC_2 \\
               c(A_2 - C_3) - bC_2 & aC_2 & aC_3 \\
             \end{array}
           \right)\,.
\]
This realisation depends on eight real parameters $A_1,A_2$, $B_1,B_2,B_3$ and $C_1,C_2,C_3$ in addition to nonzero constants $a,b$ and $c$.

Let us consider a particular case of the corresponding complete system of differential equations
\[
\frac{\mathrm d x_1}{ (x, \mathbf Ax)}=\frac{\mathrm d x_2}{ (x, \mathbf Bx)}=\frac{\mathrm d x_3}{ (x, \mathbf Cx)}\,.
\]
At $a=b=c= 1$, $A_1=A_2=1$ and
\[
B_3= 0\,,\qquad B_2= 2 - \frac{B_1}{2}\,,
\qquad
C_1= 0\,,\qquad C_2= 2 - \frac{C_3}{2}
\]
we have
\[
\Delta=x_1(x_1-x_2)(x_1 - x_3)(x_2 - x_3)(B_1 + C_3 - 4)\,.
\]
In this particular case the differential equations
\[
\left\{ \begin{array}{l}
\dot{x}_1=-2x_1(2x_1 - x_2 - x_3)\,,\\
\dot{x}_2=(x_2 - x_3)(x_2 - x_1)B_1-2x_1x_2 - 2x_1x_3 + 4x_2x_3\,,\\
\dot{x}_3=(x_3 - x_1)(x_3 - x_2)C_3-2x_1x_2 - 2x_1x_3 + 4x_2x_3\,,
\end{array}
\right.
\]
have two first integrals  
\[
H_1=x_1^{B_1(B_1 + C_3 - 4)}
(x_3-x_2)^{-4(B_1 - 2)}
(x_1 - x_2)^{2(B_1 - C_3)}
\]
and
\[
H_2=(x_3-x_1)^{B_1 - 2}x_1^{-(B_1+C_3- 4)}(x_1 - x_2)^{C_3 - 2}\,,
\]
which can be polynomials for special values of the parameters.

So, according to Monge’s viewpoint, which Lie accepted in \cite{lie-book, lie-book2},  two simultaneous equations $H_1 = c_1$ and  $H_2=c_2$ define a complete solution  depending on two constants $c_{1,2}$.

In the second example we start with Euler's vector field $e_1$ and a vector field commuting with $e_1$ and $e_2$ 
\[
e_3=x_1 (a\partial_1+b\partial_2+c\partial_3)-\frac{2a^2}{9c}x_3\partial_1\,,\qquad abc\neq 0\,.
\]
As above, we only consider the particular solution
\[
e_2= (x,\hat{\mathbf A}x)\partial_1+(x,\hat{\mathbf B}x)\partial_2+(x,\hat{\mathbf C}x)\partial_3\,,
\]
which yields a complete system of differential equations with second-order polynomial right-hand sides
\[\dot{x}_1=(x,\hat{\mathbf A}x)\,,\qquad \dot{x}_2=
(x,\hat{\mathbf B}x)\,,\qquad \dot{x}_3=(x,\hat{\mathbf C}x)\,.
\]
Here
\begin{align*}
\hat{\mathbf A}=&
\left( \begin{array}{ccc}
-bB_1 - c B_2 & \frac{ 2 a(bB_1 + cB_2) -3 bcA_1}{3b}&  A_1\\
\frac{ 2 a(bB_1 + cB_2) -3 bcA_1}{3b}&  0&  -\frac{2a^2 B_1}{9c}\\
A_1&  -\frac{2a^2 B_1}{9c} &  -\frac{4 a^2 B_2}{9c}
  \end{array}
           \right)\,,\\
           \\
\hat{\mathbf B}=&
\left( \begin{array}{ccc} 
-\frac{3b(bB_1 b + cB_2)}{2a}&  bB_1&  bB_2\\
bB_1&  \frac{c(2aB_2 - 3bB_3 - 3A_1)c - abB_1}{3b}&  bB_3\\
bB_2&  bB_3&  \frac{-bc(4aB_2 + 3bB_3 - 3A_1) - ab^2 B_1}{3c^2}
\end{array}
           \right)\,,          
\end{align*}
and
\[
\hat{\mathbf C}=
\left( \begin{array}{ccc} 
-\frac{3c(bB_1 + cB_2)}{2a}&  c B_1&  c B_2\\
c B_1&  0& \frac{(2cB_2-bB_1) - 3cA_1}{3b}\\
c B_2& \frac{(2cB_2-bB_1) - 3cA_1}{3b} & 2(A_1 -aB_2)
\end{array}
           \right)\,,  
\]
In this case we have only four parameters $B_1,B_2,B_3$ and $A_1$ in addition to nonzero $a,b$ and $c$.

Let us consider a particular case of the corresponding complete system of differential equations.
At $B_1= -b^{-1}cB_2$ we have
\[
\Delta=(2ax_3 - 3cx_1)(ax_3 - 3cx_1)(cx_2-bx_3)^2
\]
up to a constant factor. In this particular case the differential equations
have only one first integral 
\[
H_1=(cx_2-bx_3)^{2aB_2}\Bigl((ax_3 - 3cx_1)(2ax_3 - 3cx_1)\Bigr)^{(A_1-aB_2 - 3bB_3)}\,.
\]
A second local or multivalued first integral can be obtained in the framework of the Euler-Jacobi theorem \cite{jac} using the last Jacobi multiplier $M=\Delta^{-1}$.

\subsection{Solvable algebra $\mathfrak r'_{3,\lambda}$}
Let us substitute vector fields 
\[
e_1=\phi_1(x)\partial_1+\phi_2(x)\partial_2+\phi_3(x)\partial_3\,,
\quad
e_2=b\,x_2^\kappa\,\partial_2\,,
\quad e_3=a_1x_1^\kappa\partial_1+a_2x_2^\kappa\partial_2+ f(x_3)\partial_3
\]
into the  following Lie brackets
\begin{equation}\label{tp3l}
[e_1,e_2]=\lambda e_2-e_3\,,\qquad [e_1,e_3]=e_2+\lambda e_3\,,\qquad [e_2,e_3]=0\,.
\end{equation}
In this case we obtain the following complete system of differential equations 
\[
\dot{x}_1=\frac{x_1}{\kappa-1}\left(\lambda-\frac{a_1x_1^{\kappa-1}x_2^{1-\kappa}-a_2}{b}  \right)\,,\qquad
\dot{x}_2=\frac{x_2}{\kappa-1}\left(\lambda-\frac{a_2}{b}+\frac{(a_2^2+b^2)\,x_1^{1-\kappa}x_2^{\kappa-1}}{a_1b}  \right)
\]
and
\[
\dot{x}_3=-\frac{x_2^{1 - \kappa}f(x_3)}{b(\kappa-1)}+f(x_3)\left(
-\left(\lambda + \frac{a_2}{b}\right)\int \frac{\mathrm d x_3}{f(x_3)}+G(D)
\right)
\]
where $G$ is an arbitrary function on 
\[D=x_1^{1-\kappa}+a_1(\kappa-1)\,\int \frac{\mathrm d x_3}{f(x_3)}
\]

Another realisation of the vector fields $e_{2,3}$
\[
e_2=b\,x_2^\kappa\,\partial_2\,,
\qquad e_3=ax_1^\kappa\partial_1+ f(x_3)\partial_3
\]
yields another complete system of differential equations
\[
\dot{x}_1=\frac{x_1}{\kappa-1}\left(\lambda-\frac{a x_1^{\kappa-1}x_2^{1 - \kappa}}{b} +c\,x_1^{\kappa-1}\right)\,,\qquad
\dot{x}_2=\frac{x_2}{\kappa-1}\left(\lambda+\frac{bx_1^{1-\kappa}x_2^{\kappa-1}}{a}+c\,x_2^{\kappa-1}\right)\,,
\]
and
\[
\dot{x}_3=-\frac{x_2^{1 - \kappa}f(x_3)}{b(\kappa-1)}+
f(x_3)\left(-\lambda\,\int \frac{\mathrm d x_3}{f(x_3)}+G(D)
\right)
\]
where $c\in\mathbb R$ is an arbitrary number and
\[
D=x_1^{1-\kappa}+a(\kappa-1)\int \frac{\mathrm d x_3}{f(x_3)}\,.
\]
In both cases, we can choose a function $f(x_3)$ such that the resulting vector fields are globally defined.

These examples are presented only to demonstrate the existence of complete systems of differential equations associated with the solvable Lie algebra $\mathfrak r'_{3,\lambda}$ (\ref{tp3l}).

So, in this section we presented examples of integrable or complete systems in $\mathbb R^3$ which have three, two, one, or no first integrals associated with integrable groups of infinitesimal transformations in Lie's terminology. 

\section{Three-dimensional simple real Lie algebras}
\setcounter{equation}{0}

Suppose $\mathfrak g$ is a simple 3D Lie algebra over $\mathbb R$. Then $\mathfrak g$ is isomorphic
to one and only one of the following Lie algebras $\mathfrak{sl}(2,\mathbb R)$ or $\mathfrak{su}(2)$ with the brackets
\[
[e_1, e_2]=e_1\,,\qquad [ e_1, e_3]=2 e_2\,,\qquad [ e_2,e_3]=e_3\,
\] 
and
\[
[e_1,e_2]=e_3\,,\qquad [e_3,e_1]=e_2\,,\qquad [e_2,e_3]=e_1
\]
respectively \cite{gov}.

Following \cite{ts26} we consider three vector fields 
\begin{equation}\label{efh-def}
 E=\phi_1(x)\partial_1+\phi_2(x)\partial_2+\phi_3(x)\partial_3\,,\qquad  H=2(x_1\partial_1+x_2\partial_2+x_3\partial_3)
\qquad
 F=a_1\partial_1+a_2\partial_2+a_3\partial_3\,,
\end{equation}
which form the Cartan-Weyl basis  of the Lie algebra $\mathfrak{sl}(2)$
\begin{equation}\label{sl2}
[ H, E]= 2 E\,,\qquad [ H, F]=-2 F\,,\qquad [ E, F]=  H\,.
\end{equation}
Here $H$ is the Cartan operator, which measures “weight” (eigenvalues) in representations, whereas  $E$ and $F$ are the raising (creation) and lowering (annihilation) operators and $a_i\in\mathbb R$ are parameters of this realisation.

Partial differential equations $F(f)=0$ and $H(f)=0$ are completely integrable since characteristic functions of the constant vector field  $ F$ are functions of linear polynomials 
\[
d_{ij}=a_ix_j-a_jx_i\,,\qquad F(d_{ij})=0\,, \qquad F\wedge  H=\sum _{i,j=1}^3 d_{ij}\partial_i\wedge\partial_j\,,
\]
whereas characteristic functions of the vector field $H$ are homogeneous functions of zero degree. 

\begin{prp}
If  $a_1a_2a_3\neq 0$ and  $\Delta\neq 0$, then solutions of PDEs associated with brackets (\ref{sl2}) are
\begin{equation*} 
  \phi_1(x)=-\frac{x_1^2}{a_1} +g_1(x)\,,\qquad
  \phi_2(x)=-\frac{x_2^2}{a_2} +g_2(x)\,,\qquad
  \phi_3(x)=-\frac{x_3^2}{a_3}+ g_3(x)\,,
\end{equation*} 
where the functions $g_j(x)$ satisfy
\[
 (x_1\partial_1+x_2\partial_2+x_3\partial_3)g_j=2g_j\qquad\mbox{and}\qquad  F(g_j)=0\,,\qquad j=1,2,3\,.
\]
Thus the functions $g_j(x)$ are arbitrary smooth homogeneous functions of degree  2 depending only on the invariants 
\[d_{ij}=a_ix_j-a_jx_i\,,\]
of the constant vector field $F$ (\ref{efh-def}).

As an example they could be homogeneous polynomials of the second order in $x_1,x_2$ and $x_3$
\[
  g_k(x)=\sum \alpha_{k}^{ij\ell m} d_{ij}d_{\ell m}\,,\qquad \alpha_{k}^{ij\ell m}\in\mathbb R\,,
\]
depending on parameters $\alpha_{k}^{ij\ell m}$.
\end{prp}
The proof consists of directly solving the corresponding partial differential equations. We do not consider here other solutions when $a_1=0$ or $a_1=a_2=0$, or $\Delta=0$.

According to the Jacobi-Clebsch-Deahna-Frobenius-Lie theorem we have a system of completely integrable differential equations of the form
\[
\dot{x}_k=\phi_k\equiv -\frac{x_k^2}{a_k}+g_k(x)\,,\qquad k=1,2,3.
\]
The corresponding flow preserves Lie's determinant
\[
\Delta=\left\| 
\begin{array}{ccc}
x_1&x_2 &x_3\\
\phi_1&\phi_2&\phi_3\\
a_1& a_2&a_3
\end{array}
\right\|\neq 0
\]
and Jacobi's multiplier $M=\Delta^{-1}$ at $\Delta\neq 0$.

In a particular case we obtain the so-called Darboux-Halphen system 
\begin{equation}\label{halp-sys}
\left\{ \begin{array}{l}
\dot{x}_1=x_2x_3-x_1(x_2 + x_3)\,,\\
\dot{x}_2=x_1x_3-x_2(x_1+x_3)\,,\\
\dot{x}_3=x_1x_2-x_3(x_1+x_2)\,.
\end{array}
\right.
\end{equation}
It first appeared in Darboux’s work on triply orthogonal surfaces \cite{darb-3}. The general solution to (\ref{halp-sys}) was determined in 1881 by  Halphen \cite{hal} and Brioschi \cite{bri}  in terms of the elliptic modular function.

Using these complete solutions we can get first integrals
\begin{align}
f_1=&\frac{x_2}{\sqrt{x_2 - x_3}}\,\mathsf{K}\left(z\right)
 - \sqrt{x_2 - x_3}\,\mathsf{E}\left(z\right)\,, \qquad z=\frac{\sqrt{x_1 - x_3}}{\sqrt{x_2 - x_3}}
\nonumber\\
\label{f12-dh}\\
f_2=&\frac{x_3}{\sqrt{x_2 - x_3}}\,\mathsf{K}'\left(z\right)
 + \sqrt{x_2 - x_3}\,\mathsf{E}'\left(z\right)\,,
 \nonumber
\end{align}
Here $\mathsf{K}(z)$ and $\mathsf{K}'(z)$ are complete and complementary complete elliptic integrals of the first kind, whereas $\mathsf{E}(z)$ and $\mathsf{E}'(z)$ are complete and complementary complete elliptic integrals of the second kind. So, similar to Euler's examples in \cite{eul},  $f_{1,2}$ are multivalued transcendental functions. 

The physical contexts in which  equations (\ref{halp-sys}) have appeared include the dynamics of pairs of magnetic monopoles, the theory of vacuum Einstein equations for hyperk\"{a}hler Bianchi IX metrics, similarity reductions of associativity equations on a three-dimensional Frobenius manifold.

The mathematical contexts in which  equations (\ref{halp-sys}) have appeared include systems of orthogonal coordinates, Kowalevski-Painlev\'{e} meromorphic integrability,  Ramanujan's vector field on the ring of quasimodular forms, etc.

In this note, we present the Darboux-Halphen system (\ref{halp-sys}) to demonstrate the difference between the two constructions of completely integrable systems by Lie. To produce this system within the framework of the theory of groups of infinitesimal transformations, we can use globally defined vector fields. To produce the same system in the framework of function group theory, we have to use multivalued characteristic functions $f_{1,2}$ (\ref{f12-dh}). Nevertheless, both methods allow us to get various completely integrable generalisations of Darboux-Halphen equations (\ref{halp-sys}), see details \cite{ts26}.

\section{Integration of the Euler equations using $\mathfrak{aff}(\mathbb R)\times \mathbb R$ algebra}
Let us consider the classical Euler equations on $\mathfrak{so}^*(3)$
\begin{equation}\label{e-top}
\dot{x}_1=a_1x_2x_3\,,\qquad \dot{x}_2=a_2x_1x_3\,,\qquad \dot{x}_3=a_3x_1x_2\,,\qquad a_k\in\mathbb R\,.
\end{equation}
In \cite{eul-rb} Euler  showed that these differential equations  have a complete solution expressed via an elliptic integral, while Jacobi's solution expresses the same complete solution  in terms of Jacobi's elliptic functions \cite{jac50}.

Our aim is to reproduce Euler's solution in  the framework of Lie's theory. Using first integrals
\[
d_1 = x_2^2a_3 - x_3^2a_2\,,\qquad
d_2 = x_1^2a_3 - x_3^2a_1\,,\qquad
d_3 = x_1^2a_2 - x_2^2a_1\,,
\]
and  Pfaff's theorem \cite{pfaff}, we introduce three independent variables
\[
y_1=d_1\,,\qquad y_2=d_2\,,\qquad y_3=\alpha(x_1,x_2,x_3)
\]
in which the original equation
\[
\frac{d x_1}{a_1x_2x_3}=\frac{d x_2}{a_2x_1x_3}=\frac{d x_3}{a_3x_1x_2}=dt
\]
has the following form
\[
\frac{d y_1}{0}=\frac{d y_2}{0}=\frac{d y_3}{Y_3}=dt\,,
\]
where we use Jacobi's standard notation \cite{jac}. So, we reduce the solution of the original equation to a single quadrature
\[
\int \frac{d y_3}{Y_3(y_1,y_2,y_3)}=t+C_3\,,\qquad C_3\in \mathbb R\,,
\]
 involving some unknown functions $y$ and $Y$ on $x$ and constants $C_{1,2}=y_{1,2}$. Such a quadrature was explicitly obtained by Euler using kinetic energy and angular momentum \cite{eul-rb}.  

Using Lie's “new interpretation of complete solutions of differential systems” we introduce
Euler's vector field
\[
e_1=x_1\partial_1+x_2\partial_2+x_3\partial_3
\]
and the vector field associated with differential equations (\ref{e-top})
\[
e_2=a_1x_2x_3\partial_1+a_2x_1x_3\partial_2+a_3x_1x_2\partial_3\,,
\]
so that
\[ [e_1,e_2]=e_2\,.\]
Let us introduce a third vector field depending on a function $\alpha(x_1,x_2,x_3)$
\[e_3=\alpha(x_1,x_2,x_3) e_2\,,\]
which satisfies the equations
\begin{equation}\label{v-eul}
[e_1,e_3]=0\,,\qquad [e_2,e_3]=\lambda e_2\,.
\end{equation}
It means that 
\[
e_2(\alpha)=\lambda\,,\qquad \Rightarrow\qquad \frac{\mathrm d}{\mathrm d t}\, \alpha(x_1,x_2,x_3)=\lambda\,,
\]
and after transformation $e_3\to e_1 + \lambda^{-1}e_3$ we obtain (\ref{t3l}), i.e. 
we consider a solvable Lie algebra isomorphic to $\mathfrak{aff}(\mathbb R)\times \mathbb R$.

Brackets (\ref{v-eul}) yield a system of partial differential equations which can be solved using modern computer algebra systems in a few seconds. Let us present one of the solutions
\[
\alpha(x_1,x_2,x_3)=-\frac{\mathrm{i}\lambda}{\sqrt{a_1d_1}}\,{\mathsf F}
\left(\frac{x_1}{x_2}\,\sqrt{\frac{d_1}{d_2}}, \sqrt{\frac{a_2d_2}{a_1d_1}}\right)\,.
\]   
Here ${\mathsf F}(z,\kappa)$ is the incomplete elliptic integral of the first kind which is identical to the inverse of Jacobi's $sn(z,\kappa)$ function. 

Let us note that projective geometry was second nature to Lie and, therefore, the appearance of a projective coordinate in the quadrature  obtained using Lie's method is not at all surprising
\[
\frac{x_1}{x_2}=\sqrt{\frac{a_1x_3^2+d_2}{a_2x_3^2+d_1}},\qquad d_{1,2}=C_{1,2}\,.
\]
This example demonstrates that from the knowledge that the differential equations admit a group of known transformations, one is able to obtain information about its integration.

\section{Conclusion}
The following formulation of the integrability theorem attributed to Lie can be found in modern literature  \cite{grab,grab2,koz}:\\
\textit{"If $n$ vector fields, $X_1,\ldots,X_n$, which are linearly independent at each point
of an open set $U\subseteq\mathbb R^n$, span a solvable Lie algebra and satisfy
$[X_1, X_i ] =\lambda_i X_1$ with $\lambda_i \in\mathbb R$, then $X_1$ is integrable by quadratures in $U$."} 
\\
 We only suspect that the authors in \cite{grab, grab2, koz} simply shortened the full sentence "integrable by quadratures using the Lie reduction method" when interpreting Lie's ideas and forgot that Lie always assumes the existence of a non-zero Jacobian (invariant measure) describing the change of variables according to the ideas of Pfaff and Jacobi.

In this note we briefly recall some Lie results on noncommutative integrability both in terms of infinitesimal transformations and in terms of characteristic functions which give rise to the Lie function group. We demonstrate that
for some completely integrable systems, first integrals are defined globally, whereas vector fields are only defined locally and it is necessary to use Lie's theory of characteristic strips. Conversely, for other systems, vector fields are defined globally while first integrals are defined only locally. Of course, there is also a third class of integrable systems having only local first integrals and local  vector fields, since the definition of complete integrability has a local origin.

\vskip0.2truecm
The research  was carried out with the financial support of the Ministry of Science and Higher Education of the Russian Federation in the framework of a scientific project under agreement No 075-15-2025-013 by St. Petersburg State University as part of the national project “Science and Universities” in 2025.
\vskip0.2truecm
The author gratefully acknowledges the kind hospitality provided by
Yanqi Lake Beijing Institute of Mathematical Sciences and Applications during his stay in Fall 2025 when work on this text was finished.


\begin{thebibliography}{99}
\bibitem{im69}
Imschenetsky V., Sur l'intégration des équations aux dérivées partielles du premier ordre, \newblock{\em Archiv für Math. und Physik}, 50 (1869), 278-474.

\bibitem{lie-book}
Lie S., \newblock{\em Theorie der Transformationsgruppen. Abhandlung I}, unter Mitwirkung von Dr. Friedrich Engel, Leipzig, 1888.

\bibitem{lie-book2}
Lie S., Scheffers G., \newblock{\em Vorlesungen \"{u}ber continuierliche Gruppen mit geometrischen und anderen Anwendungen}, Leipzig, B G Teubner, 1893.

\bibitem{for90}
Forsyth A.R., \newblock{\em Theory of Differential Equations}, vol 1-6, Cambridge University
Press (1890-1906), reprinted by Dover Public., New York (1959).


\bibitem{gour91}
Goursat E., \newblock{\em 
Leçons sur l'intégration des équations aux dérivées partielles du premier ordre}, Paris, Hermann Librairie Scientifique, 1891.

\bibitem{car}
Cartan E., \newblock{\em Le\c{c}ons sur les invariants int\'{e}graux}, Hermann, Paris, 1922. 

\bibitem{jac-27}
Jacobi C.G.J.,  \"{U}ber die Integration der partiellen Differentialgleichungen erster Ordnung.
 Journal f\"{u}r die reine und angewandte Mathematik, 2 (1827), 317–329.

\bibitem{olver}
Olver P. J., \newblock{\em  Applications of Lie Groups to Differential Equations}, 2nd ed., Grad. Texts in Math., vol. 107, New York: Springer, 1993.

\bibitem{lag81}
Lagrange J.L., Sur diff\'{e}rentes questions d’analyse relatives \`{a} la th\'{e}orie des integrales particuli\`{e}res, \newblock{\em Nouv. M\'{e}m. Acad. Sci. Berlin}, (1781), Reprinted in {\OE}euvres compl\`{e}tes tome 4, 585–634.

\bibitem{lag85}
Lagrange J.L., 
M\'{e}thode g\'{e}n\'{e}rale pour int\'{e}grer les \'{e}quations aux diff\'{e}rences partielles du premier ordre, lorsque ces diff\'{e}rences ne sont que lin\'{e}aires, \newblock{\em Nouv. M\'{e}m. Acad. Sci. Berlin}, (1785),
Reprinted in {\OE}uvres compl\`{e}tes, tome 5, 543-562.

\bibitem{pfaff}
Pfaff J. F., Methodus generalis aequationes differentiarum partialium, nec non aequationes
differentiales vulgares, utrasque primi ordinis, inter quotcunque variabiles, complete integrandi,
Abhandlungen d. Akad. der Wiss. zu Berlin 1814–15 (1818), 76-136.

\bibitem{lie-00}
Lie S., Zur Theorie der Differentialprobleme, Forhandlinger Christiania 1872,
(1873), 132-133. Reprinted in Abhandlungen 3, 27-28.  

\bibitem{lie-1873a}
Lie S., Partielle Differentialgleichungen erster Ordnung, in denen die unbekannte
Funktion explizite vorkommt,
\newblock{\em  Forhandlinger Christiania 1873}, (1874), 52-85. Reprinted in Abhandlungen 3, 64-95.

\bibitem{lie-1873c}
Lie S., \"{U}ber partielle Differentialgleichungen erster Ordnung,
\newblock{\em  Forhandlinger
Christiania 1873}, (1874), 16-51. Reprinted in Abhandlungen 3, 32-63. 

\bibitem{lie-0}
Lie S., Zur Theorie des Integrabilit\"{a}tsfaktors,
\newblock{\em Forhandlinger Christiania 1874}, (1875), 242-254. Reprinted in Abhandlungen 3, 176-186. 

\bibitem{lie-1}
Lie S., Verallgemeinerung und neue Verwertung der Jacobischen Multiplikatortheorie, 
\newblock{\em Forhandlinger Christiania 1874}, (1875), 255-274. Reprinted in Abhandlungen 3, 188-205. 


\bibitem{lie-1874b}
Lie S., Begr\"{u}ndung einer Invariantentheorie der Ber\"{u}hrungstransformationen,
\newblock{\em Math. Ann.}, 8 (1874), 215-303. Reprinted in Abhandlungen 4, 1-96.

\bibitem{lie77}
Lie S., Theorie des Pfaffschen Problems I, \newblock{\em  Archiv for Mathematik}, 2 (1877), 338-379. Reprinted
in Abhandlungen 3, 320-351.

\bibitem{engel}
Engel F., Anmerkungen, in \newblock{\em Sophus Lie Gesammelte Abhandlungen}, 3,  585–789,
Teubner, Leipzig, 1922.

\bibitem{frob}
Frobenius G. \"{U}ber das Pfaffsche Problem,
\newblock{\em Journal f\"{u}r die reine und angewandte Mathematik}, 82 (1877), 230-315.

\bibitem{sch80}
Scholz E., \newblock{\em Geschichte des Mannigfaltigkeitsbegriffes von Riemann bis Poincar\'{e}},
Boston, Birkh\"{a}user, 1980.

\bibitem{lie-1884a}
Lie, S., Allgemeine Untersuchungen \"{u}ber Differentialgleichungen, die eine kontinuierliche,
endliche Gruppe gestatten, \newblock{\em Math. Ann.}, 25 (1885), 71-151. Reprinted
in Abhandlungen 6, 139-223.


\bibitem{st1}
W. Stekloff,  Application d'un th\'{e}or\`{e}me g\'{e}n\'{e}ralis\'{e} de Jacobi
au probl\`{e}me de S.\,Lie--Mayer,
\newblock{\em Comptes rendus des s\'{e}ances de l'acad\'{e}mie des sciences}, 148 (1909), 277-279.

\bibitem{st2}
W. Stekloff, Application du th\'{e}or\`{e}me g\'{e}n\'{e}ralis\'{e} de Jacobi au probl\`{e}me de
Jacobi--Lie, \newblock{\em Comptes rendus des s\'{e}ances de l'acad\'{e}mie des sciences}, 148 (1909),
465-468.

\bibitem{fom}
Mishchenko A.S., Fomenko A.T.,  Generalized Liouville method of integration of
Hamiltonian systems, Funct. Anal. Appl., 12 (1978), 113-121.

\bibitem{jac}
Jacobi C.G.J., Theoria novi multiplicatoris systemati {\ae}quationum differentialium vulgarium
applicandi,  \newblock{\em Journal f\"{u}r die reine und angewandte Mathematik}, 27 (1844), 199-268, 29 (1845), 213-279 and 333-376.

\bibitem{koz}
 Kozlov V.V., Solvable algebras and integrable systems, \newblock{\em  Regul. Chaotic Dyn.}, 29:5 (2024),
717–727.

\bibitem{gov}
Gorbatsevich V., Onishchik A.,  Vinberg E.,
\newblock{\em  Structure of Lie groups and
Lie algebras}, English transl. in Encycl. Math Sc. 41, Springer-Verlag,
Berlin, Heidelberg, 1994.

\bibitem{eul}
Euler L., \newblock{\em Institutionum calculi integralis volumen primum}, Petropoli: Acad. Imp. Sci., 1768.

\bibitem{sw1}
Strelcyn J.-M., Wojciechowski S.,
A method of finding integrals for three-dimensional dynamical systems,
\newblock{\em Physics Letters A} 133 (1988), 207-212.

\bibitem{sw2}
Grammaticos B., Ollagnier J.M.,  Ramani A.,  Strelcyn J.-M., Wojciechowski S.,
Integrals of quadratic ordinary differential equations in $\mathbb R^3$: the
Lotka-Volterra system, \newblock{\em Physica A}, 163 (1990), 683-722.

\bibitem{os90}
Ollagnier J.M., Strelcyn J-M.,
\newblock{\em On first integrals of linear systems, Frobenius integrability theorem and linear representations of Lie algebras}. In: Fran\c{c}oise, J-P., Roussarie, R. (eds) Bifurcations of Planar Vector Fields, Lecture Notes in Mathematics, vol 1455. Springer, Berlin, Heidelberg, (1990).

\bibitem{ts26}
Tsiganov A.V., On the Darboux-Halphen system: Jacobi vs Lie, 
\newblock{\em J. Geom. Phys.}, 228 (2026), 105914.

\bibitem{darb-3}
Darboux G., Sur la th\'{e}orie des coordonn\'{e}es curvilignes et les syst\`{e}mes orthogonaux,
\newblock{\em  Ann. Ec. Norm. Sup\'{e}r.}, 7 (1878), 101-150.

\bibitem{hal}
Halphen G. H., Sur un syst\`{e}me d’\'{e}quations diff\'{e}rentielles, 
\newblock{\em C. R. Acad. Sci. Paris}, 92 (1881), 1101-1103.

\bibitem{bri}
Brioschi F., Sur un syst\`{e}me d’\'{e}quations diff\'{e}rentielles, 
\newblock{\em C. R. Acad. Sci. Paris}, 92 (1881), 1389-1393.

\bibitem{eul-rb}
Euler L., Nova methodus motum corporum rigidorum determinandi, \newblock{\em
Novi Commentarii academiae scientiarum Petropolitanae}, 20 (1776), 208-238.

\bibitem{jac50}
Jacobi C. G. J., 
Sur la rotation d’un corps, extrait d’une lettre adress\'{e}e
\`{a} l’Acad\'{e}mie des Sciences,
\newblock{\em Journal de math\'{e}matiques pures et appliqu\'{e}es 1re s\'{e}rie}, 14 (1849), 337-344.



\bibitem{grab}
Cariñena J.F., Falceto F.,  Grabowski J.,  Rañada M.F., 
Geometry of Lie integrability by quadratures, 
Journal of Physics A: Mathematical and Theoretical,  48:21 (2015), 215206.

 \bibitem{grab2}
Cariñena J.F., Falceto F.,  Grabowski J.,
Solvability of a Lie algebra of vector fields implies their integrability by quadratures,
Journal of Physics A: Mathematical and Theoretical, 49:42 (2016), 425202.



\end{thebibliography}
\end{document}